\documentclass[fleqn,10pt]{wlscirep}
\title{Lower bound for the spatial extent of localized modes in photonic-crystal waveguides with small random imperfections}

\author[1]{R\'emi Faggiani}
\author[2,3,*]{Alexandre Baron}
\author[1]{Xiaorun Zang}
\author[4]{Lo\"ic Lalouat}
\author[5,6]{Sebastian A. Schulz}
\author[5,7]{Bryan O'Regan}
\author[1,$\dag$]{Kevin Vynck}
\author[4]{Beno\^it Cluzel}
\author[4]{Fr\'ed\'erique de Fornel}
\author[5,7]{Thomas F. Krauss}
\author[1,2,$\ddag$]{Philippe Lalanne}
\affil[1]{LP2N, UMR 5298, CNRS – IOGS - Univ. Bordeaux, 33400 Talence, France}
\affil[2]{LCF, UMR 8501, CNRS - IOGS - Univ. Paris-Sud, 91127 Palaiseau, France}
\affil[3]{CRPP, UPR 8641, CNRS - Univ. Bordeaux, 33600 Pessac, France}
\affil[4]{ICB, UMR 6303, CNRS - Univ. Bourgogne, 21078 Dijon, France}
\affil[5]{SUPA, School of Physics \& Astronomy, University of St Andrews, St Andrews, KY16 9SS, UK}
\affil[6]{Department of Physics and Max-Planck Centre for Extreme and Quantum Photonics, University of Ottawa, Ottawa, K1N 6N5, Ontario, Canada}
\affil[7]{Department of Physics, University of York, York, YO10 5DD, UK}

\affil[*]{alexandre.baron@u-bordeaux.fr}
\affil[$\dag$]{kevin.vynck@institutoptique.fr}
\affil[$\ddag$]{philippe.lalanne@institutoptique.fr}

\begin{abstract}
Light localization due to random imperfections in periodic media is paramount in photonics research. The group index is known to be a key parameter for localization near photonic band edges, since small group velocities reinforce light interaction with imperfections. Here, we show that the size of the smallest localized mode that is formed at the band edge of a one-dimensional periodic medium is driven instead by the effective photon mass, i.e. the flatness of the dispersion curve. Our theoretical prediction is supported by numerical simulations, which reveal that photonic-crystal waveguides can exhibit surprisingly small localized modes, much smaller than those observed in Bragg stacks thanks to their larger effective photon mass. This possibility is demonstrated experimentally with a photonic-crystal waveguide fabricated without any intentional disorder, for which near-field measurements allow us to distinctly observe a wavelength-scale localized mode despite the smallness ($\sim 1/1000$ of a wavelength) of the fabrication imperfections.
\end{abstract}
\begin{document}

\flushbottom
\maketitle
%
%
\thispagestyle{empty}

\section*{Introduction}

Random imperfections, even very small ones, can have a profound impact on light propagation in periodic photonic structures, the most striking phenomenon being undoubtedly the formation of small localized modes in the vicinity of photonic band edges. The interplay between long-range order and perturbative disorder, originally proposed as a way to enable strong light localization in three-dimensional media~\cite{John1987, Conti2008}, was largely investigated in one-dimensional (1D) layered structures (i.e., Bragg stacks), which can be modeled with greater ease~\cite{McGurn1993, Bulgakov1996, Deych1998, Vinogradov2004, Kaliteevski2006, Izrailev2009, Poddubny2012}. Besides these works, the research topic arose considerable interest in the photonic-crystal community, when it was realized that the operation of slow-light devices based on photonic-crystal waveguides (PhCWs) was unavoidably limited by small residual fabrication imperfections~\cite{Notomi2001, Mookherjea2008, Mazoyer2010, Melati2014}. The possibility to observe individual localized modes formed by disorder in these structures and exploit them as ``optical cavities'' for, e.g., quantum information processing~\cite{Sapienza2010, Thyrrestrup2012, Gao2013, Minkov2013} or random lasing~\cite{Yang2011a, Liu2014} prompted numerous studies on their confinement properties~\cite{Topolancik2007, Smolka2011, Spasenovic2012, Huisman2012a}. Quite remarkably, near-field measurements on PhCWs fabricated with state-of-the-art nanotechnologies~\cite{Spasenovic2012, Huisman2012a} suggest that \textit{wavelength-scale} localized modes, comparable in size to engineered heterostructure nanocavity modes in PhCWs~\cite{Song2005, Kuramochi2006}, could naturally be formed in spite of the very low perturbation level.

It is widely accepted that the typical spatial extent of localized modes in the band decreases when approaching the edge. This trend is generally understood by drawing a parallel between the mode spatial extent and the ``Anderson'' localization length, which describes the exponential attenuation of the ensemble-averaged intensity with the system size. The localization length is known to scale as the square of the group velocity $v_g$ for small disorder levels~\cite{Hughes2005, Mazoyer2009, Garcia2010a}, thereby indicating that light confinement should be extremely strong at the band edge, where $v_g$ vanishes (in perfectly periodic media). The formation of small localized modes in close vicinity to the band edge, however, deserves special attention, as it relies not only on the interference between multiply-scattered propagating waves, leading to Anderson localization~\cite{Lagendijk2009, Segev2013}, but also on the attenuation provided by the photonic band gap. Imperfections, even vanishingly small ones, can indeed easily create gap (defect) modes, similar in nature to photonic-crystal nanocavities~\cite{Lalanne2008, Bliokh2008, Notomi2010}, which participate in the broadening of the band edge~\cite{Savona2011, Huisman2012a, Mann2015} and in the formation of the so-called Lifshitz tail in the band gap~\cite{Lifshitz1964, Huisman2012a}. In this narrow spectral range around the band edge, propagating and evanescent waves mix up and the group velocity evidently looses physical significance, thereby requiring the basic parallel between mode spatial extent and localization length to be revisited.

In this article, we theoretically, numerically and experimentally investigate the confinement properties of localized modes in close vicinity to the band edge of 1D periodic photonic structures at small disorder levels. We demonstrate in particular that the size of the smallest localized mode that may be found in a given photonic structure is driven by the \textit{effective photon mass}, i.e. the flatness of the dispersion curve, rather than the group index. This, in turn, suggests that an engineering of photonic bands in PhCWs may allow us either to lower the impact of residual imperfections on the performance of slow-light photonic devices or to reinforce the light confinement for strong light-matter interaction purposes.

The fact that the sensitivity of wave transport to random perturbations is increased with the flatness of the dispersion relation is well known in condensed matter physics~\cite{Chalker2010, Baboux2016}, but the possibility to control the spatial extent of the smaller localized modes via the effective mass has not been suggested in earlier works on light localization in photonic structures~\cite{John1987, Notomi2001, Hughes2005, Topolancik2007, Mookherjea2008, Mazoyer2009, Sapienza2010, Mazoyer2010, Garcia2010a, Smolka2011, Yang2011a, Savona2011, Thyrrestrup2012, Spasenovic2012, Huisman2012a, Gao2013, Minkov2013, Melati2014, Liu2014, Mann2015}. We believe that the conceptual understanding gained from our study could help the design of PhCWs and cavities with potential outcomes in photonic technologies.

In the remainder of this paper, we first present a phenomenological model for light propagation near the band edge of 1D periodic media with small random perturbations. We predict that a minimum number of periods is necessary to form a localized mode at a given disorder level and derive a closed-form expression relating this lower bound to the disorder level and the effective photon mass. Then, we test and validate our predictions by a series of numerical simulations on randomly-perturbed Bragg stacks and PhCWs. At tiny disorder levels, PhCWs are found to support surprisingly small localized modes, much smaller than those observed in Bragg stacks thanks to their larger effective photon mass. Finally, we verify this possibility by performing near-field measurements on a photonic-crystal waveguide fabricated without any intentional disorder and observe very distinctly a localized mode with a spatial extent of only 6 $\mu$m, in agreement with our numerical simulations.

\section*{Results}

\subsection*{Formation of localized modes at band edges}\label{sec:model}

\begin{figure}[t]
\centering
\includegraphics[width=\linewidth]{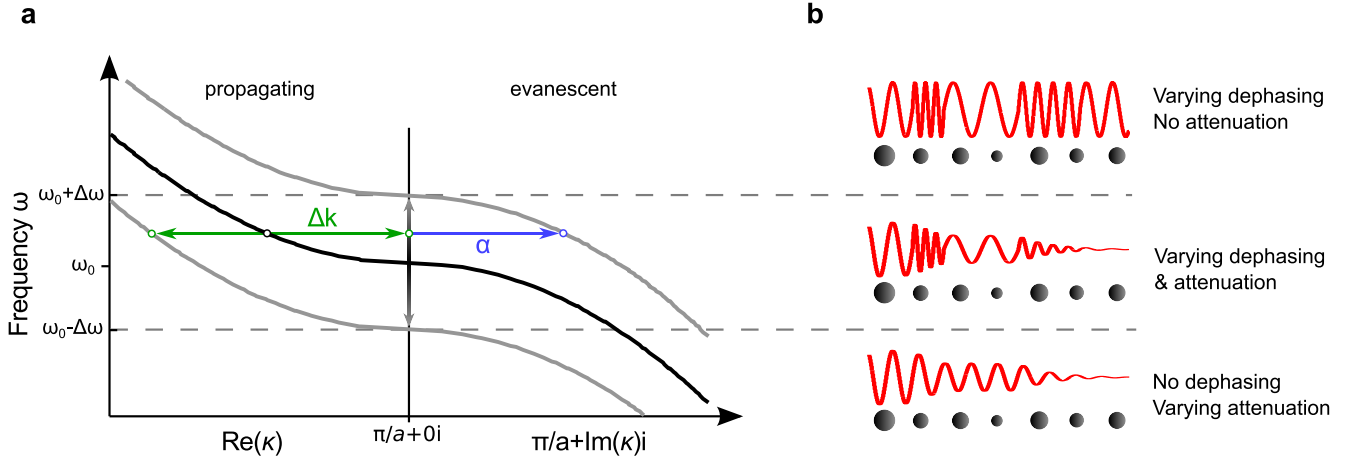}
\caption{\textbf{Effect of random imperfections on light propagation in one-dimensional periodic media.} \textbf{(a)} Sketch of the dispersion curve of a one-dimensional periodic medium near a band edge at $\omega=\omega_0$ (black curve). Note that the left and right parts of the plot correspond to the real and imaginary parts of the wavevector $\kappa$. Perturbations result in positive or negative energy shifts $\pm \Delta \omega$ of the dispersion curve (gray curves), where $\Delta \omega \ll \omega_0$, resulting in a phase-shift $\Delta k$ (green arrows) and/or a damping at rate $\alpha$ (blue arrow) for a guided wave at $\omega=\omega_0+\delta\omega$. \textbf{(b)} Sketches of wave propagation in randomly-perturbed periodic media. In the vicinity of the band edge, $\omega_0 -\Delta \omega \leq \omega \leq \omega_0+ \Delta \omega $, light successively and randomly experiences either phase-shifting or damping unit cells (middle), contrary to higher and lower frequencies, for which only phase-shifts or tunnelling damping are experienced (top and bottom, respectively).}
\label{fig:1}
\end{figure}

We start by considering an arbitrary one-dimensional periodic photonic structure and aim at establishing an explicit relation between the dispersion curve of the unperturbed medium, the level of geometric variation, and the spatial extent of the resulting localized modes at the band edge. A typical dispersion curve $\omega(\kappa)$ near a band edge is sketched in Fig.~\ref{fig:1}(a). While it is common practice in the literature to plot band structures for purely real wavevectors only, one should be aware that the wavevector is in general a complex quantity, $\kappa=k+i\alpha$. Dispersion relations are analytic and continuous at band edges~\cite{Kohn1959} and can be approximated by a quadratic expression
\begin{equation}\label{eq:dispersion}
\omega - \omega_0 = \frac{(\kappa -\pi/a)^2}{2m},
\end{equation}
where $m=(\partial^2 \omega / \partial \kappa^2)^{-1}$ is the effective photon mass, which describes the flatness of the dispersion curve. The complex wavevector equals $\kappa=k$ in the band, corresponding to propagating waves, and $\kappa=\pi/a + i \alpha$ in the gap, leading to an exponentially-damped evanescent wave. Note that Eq~(\ref{eq:dispersion}), which comes from a Taylor expansion of the dispersion relation at the band edge, remains valid for complex periodic waveguides provided that the mode dispersion remains below the light line of the cladding. It is hence often used in photonics to describe band-edge phenomena~\cite{Song2005,Povinelli2005,Xue2016} (albeit for real wavevectors only in most cases).

As sketched in Fig.~\ref{fig:1}(a), small geometrical variations result in energy shifts $\pm \Delta \omega$ of the dispersion curve with negligible deformation~\cite{Soljacic2004}. In the perturbative regime, the relative frequency shift $\Delta\omega/\omega$ is directly proportional to the variation of the waveguide effective refractive index $\Delta n_\text{eff}/n_\text{eff}$, which itself scales linearly with the disorder level $\sigma$. These energy shifts define three frequency regions, where different behaviors are expected. At frequencies $\omega < \omega_0 - \Delta \omega$, Bloch waves are essentially evanescent (bottom panel in Fig.~\ref{fig:1}(b)). In this regime, light is expected to be strongly damped via the band gap attenuation and few gap modes may be found. By contrast, at frequencies $\omega > \omega_0 + \Delta \omega$, Bloch waves are essentially propagating (top panel in Fig.~\ref{fig:1}(b)). This is typically the regime where Anderson-localized modes are found and the localization length should scale as $\xi \sim 1/n_g^2$. To understand the limited range of validity of this regime, one should know that this scaling behavior comes from a double limit in the group velocity $v_g=c/n_g$ and the disorder level $\sigma \propto \Delta \omega$ both tending towards zero. It is valid only when $\sigma$ tends towards zero at a fast enough rate compared to $v_g$. This requirement guarantees that the impact of random imperfections on transport remains perturbative~\cite{Wang2008}. Due to the nonvanishing disorder level in real nanostructures, this condition necessarily fails in the close vicinity of the band edge and the $1/n_g^2$ scaling behavior unavoidably breaks down. This peculiar regime that surrounds the band edge, $\omega_0 -\Delta \omega \leq \omega \leq \omega_0 + \Delta \omega $, is the one of interest in this article. Here, light propagating in the randomly-perturbed periodic medium will experience alternatively and randomly either phase-shifts $\Delta k$ (in $\text{Re}(\kappa)$) or exponential attenuations $\alpha$ (in $\text{Im}(\kappa)$). The formation of localized modes should therefore rely on an interplay between propagating and evanescent waves (middle panel in Fig.~\ref{fig:1}(b)).

We adopt a Fabry-Perot picture to model the formation of a localized mode in this regime. By analogy with standard optical cavities, for a localized mode to appear at a frequency $\omega$ (it is convenient to situate the frequency in relation to the band edge, $\omega=\omega_0+\delta\omega$), it is required that, during its transport, light both accumulates a phase-shift that satisfies a phase-matching condition (typically, $2\pi$ on a round trip) and experiences a strong damping (typically, $1/e^2$ for the intensity). This simple reasoning first indicates that \textit{a minimum number of periods} $N_\text{min}$ \textit{is necessary to form a localized mode in a perturbed medium}.

Evidently, smaller localized modes are formed for imperfections that produce larger momentum variations and damping rates. Depending on the perturbation and the frequency, the momentum variations (with respect to the unperturbed mode) can be either positive or negative, and the damping rate can be zero or take large values. To account for these statistical variations, we calculate the average momentum variation $\langle \Delta k \rangle$ and the average damping rate $\langle \alpha \rangle$ using Eq.~(\ref{eq:dispersion}) and, keeping only the lowest order in $\delta\omega/\Delta \omega$, find that
\begin{equation}
\langle \Delta k \rangle \approx
\begin{cases}
\frac{1}{2} \sqrt{2 m \Delta \omega} \left(1 + \frac{\delta\omega}{2\Delta\omega} \right) \quad &\text{for } \delta\omega \leq 0, \\
\frac{1}{2} \sqrt{2 m \Delta \omega} \left( 1 - 2 \sqrt{\frac{\delta\omega}{\Delta\omega}} \right) \quad &\text{for } \delta\omega \geq 0,
\end{cases}
\end{equation}
and
\begin{equation}
\langle \alpha \rangle \approx \frac{1}{2} \sqrt{2 m \Delta \omega} \left(1 - \frac{\delta\omega}{2\Delta\omega} \right).
\end{equation}
Hence, $\langle \Delta k \rangle$ is the largest at the band edge ($\omega=\omega_0$) and decays more slowly in the gap than in the band ($\delta\omega/2\Delta\omega$ compared to $ 2 \sqrt{\delta\omega/\Delta\omega}$), while $\langle \alpha \rangle$ continuously increases when entering deeper into the gap. This indicates, on the one hand, that \textit{the smallest localized modes should be found at the band edge} ($\omega=\omega_0$), and on the other hand, that the localized modes formed in the gap should be smaller and more numerous than in the band. To obtain the lower bound value for the mode spatial extent, it is thus sufficient to consider the momentum variation and damping rate at the band edge, $\langle \Delta k \rangle = \frac{1}{2} \sqrt{2m \Delta \omega}$ and $\langle \alpha \rangle=\frac{1}{2} \sqrt{2m \Delta \omega}$, which evidently become more important as the dispersion curve flattens, i.e. for larger effective photon masses. This leads us to predict that the lower bound on the spatial extent of localized modes in perturbed periodic media scales as
\begin{equation}\label{eq:1}
N_\text{min} \propto (a^2 m \Delta \omega)^{-1/2}.
\end{equation}
Equation~(\ref{eq:1}) is obtained by neglecting intricate multiple-scattering processes in transport, such as variations of the reflection and transmission coefficients at the perturbed lattice sites~\cite{Baron2011}, yet it is highly robust and accurate. As will be seen with numerical and experimental results, fine effects related to the actual geometry, especially the scattering coefficients at every lattice sites, impact the proportionality factor but not the scaling with $m$ and $\Delta\omega$.


\subsection*{Threshold in the size distribution of localized modes}

To test our prediction on the existence of a minimum number of periods necessary to form a localized mode near a photonic band edge, we resort to numerical simulations. A systematic exploration of real PhCWs with long-scale propagation lengths and nanometer-scale perturbations would require unreasonably long computation times with 3D fully-vectorial approaches, considering as well the large number of simulated structures to reach good statistical accuracy. Fortunately, in close vicinity of the band edge, that is the region of interest here, out-of-plane scattering into the air cladding is much weaker than backscattering (scaling as $n_g$ and $n_g^2$, respectively)~\cite{Hughes2005,OFaolain2010}. Since we are interested in the spatial extent properties of localized modes (and not Q-factor properties), we then believe that the problem can safely be treated within a 2D approximation.

\begin{figure}[t]
\centering
\includegraphics[width=\linewidth]{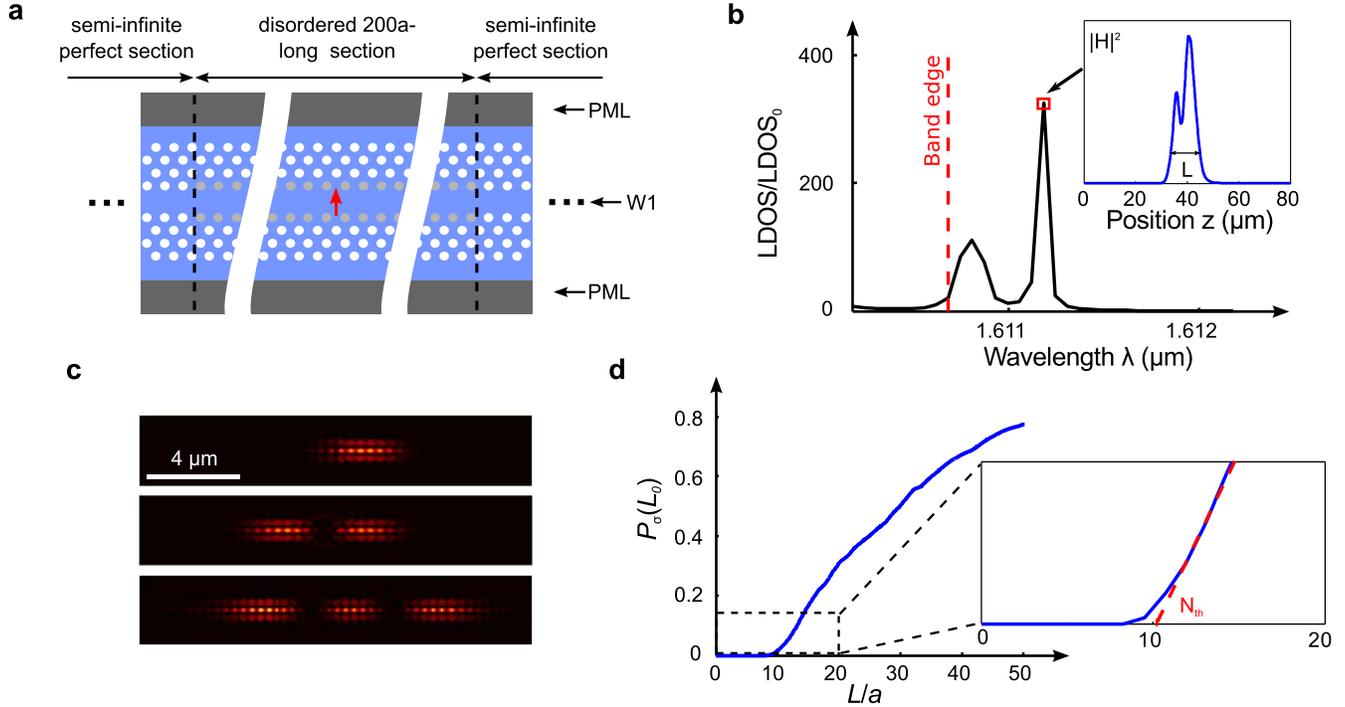}
\caption{\textbf{Numerical study of small localized modes formed in W1 waveguides.} \textbf{(a)} Computational layout used to estimate $P_\sigma(L_0)$. The red arrow represents a dipole source placed in the center of a $100a$-long section of the PhCW. The grey holes of the two inner rows represent the perturbed holes. Two semi-infinite unperturbed PhCWs surround the perturbed section. \textbf{(b)} Spectral dependence of the LDOS normalized to that in free space, LDOS$_0$, in a specific configuration. The vertical red dashed line indicates the band edge wavelength. The envelop of the magnetic-field-intensity profile $|H|^2$ of the resonant mode marked by a red square is shown in the inset. Its spatial extent is $L=12.4$ $\mu$m. \textbf{(c)} Examples of the intensity distribution $|\textbf{E}|^2$ of small localized modes obtained by numerical calculations with $\sigma = 1.5$ nm. The localized modes may be composed of several sub-spots. \textbf{(d)} Numerical prediction of $P_\sigma(L_0)$ for a W1 waveguide with $\sigma = 1.5$ nm. The distribution exhibits a clear threshold $N_\text{th}$, obtained from a linear fit (red dashed line).}
\label{fig:2}
\end{figure}

We perform 2D fully-vectorial calculations with a Fourier-Bloch-mode method~\cite{Silberstein2001, Lecamp2007b} on a single-row-missing (W1) PhCW with an hexagonal lattice constant $a=420$ nm, a hole radius $0.3a$ and an effective index of 2.83 to model the transverse confinement of the main TE-mode in a silicon membrane of thickness 240 nm suspended in air~\cite{Hugonin2007}. We implement disorder by varying the hole radii in the first rows of the W1 waveguide according to a normal distribution with standard deviation $\sigma$. The computational layout is shown in Fig.~\ref{fig:2}(a). In brief, we calculate the local density of states (LDOS) spectrum in the center of a 100$a$-long perturbed medium and spectrally locate the resonant modes of the system from the observed Lorentzian peaks, see Fig.~\ref{fig:2}(b) for an example. The LDOS spectrum is calculated on a narrow frequency window close to the band-edge wavelength $\lambda_0=2\pi c/\omega_0$ ($\lambda_0 - 0.5 \; \text{nm} < \lambda < \lambda_0 + 1.5 \; \text{nm}$), in which, according to our analysis above, the smallest possible localized modes should be observed. We apply a strict protocol, described in the Methods section, to ensure that the peaks correspond to individual localized modes, i.e. they are not affected by the finite length of the computational window. Repeating the calculation for 900 independent disorder realizations, we estimate the distribution function $P_\sigma(L_0)$, which represents the likelihood of observing a localized mode near the band edge frequency with a spatial extent $L$ smaller than $L_0$ at any position along an infinitely-long W1 waveguide perturbed by a disorder level $\sigma$.

$P_\sigma(L_0)$ is shown in Fig.~\ref{fig:2}(d) for a disorder level $\sigma = 1.5$ nm, which is comparable to the residual disorder amplitude left by state-of-the-art nanofabrication technologies. The curve clearly exhibits a threshold-like behavior and evidences that localized modes are formed essentially above a certain threshold length $N_\text{th}a$, where $N_\text{th}$ is the number of lattice periods. The threshold is straightforwardly estimated by a linear fit of $P_\sigma(L_0)$ for $0.03 < P_\sigma < 0.15$. The intensity maps $|\textbf{E}|^2$ of three localized modes are shown in Fig.~\ref{fig:2}(c), the first one corresponding to the smallest mode ($L = 3.85$ $\mu$m) we obtained among the 900 realizations, the others showing that localized modes may be composed of several sub-spots. Quite remarkably, our numerical simulations unambiguously show that wavelength-scale localized modes may be observed in PhCWs at disorder levels of the order of $\lambda/1000$. This point will be further discussed below. It is also interesting to remark that most of the smaller localized modes were formed in the gap region, as expected from the perturbative analysis in the previous section.

\subsection*{Numerical verification of the scaling law}

\begin{figure}[t]
\centering
\includegraphics[width=\linewidth]{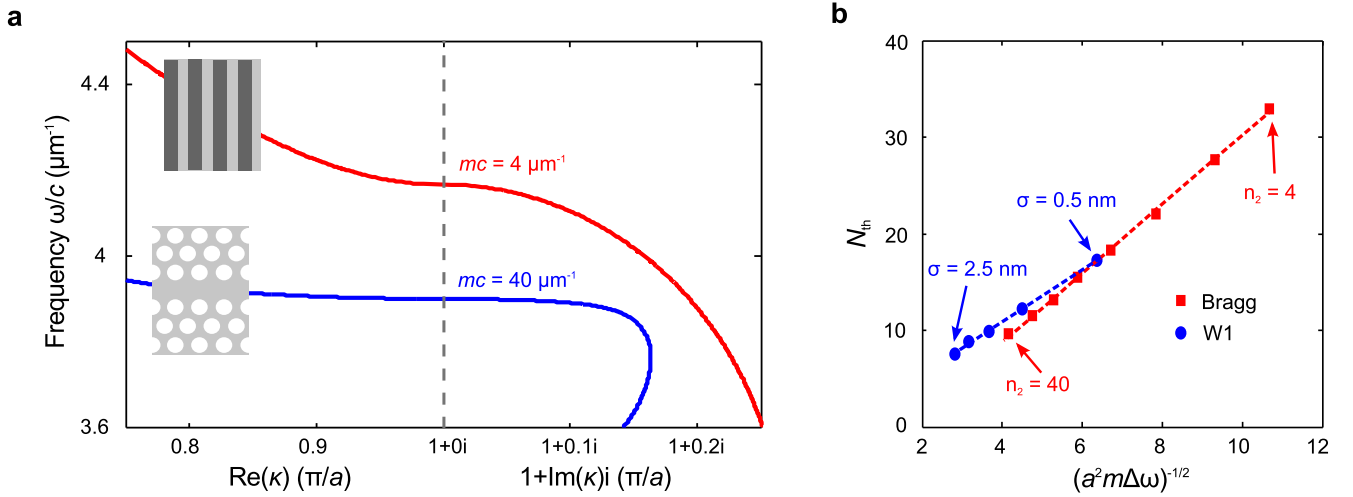}
\caption{\textbf{Scaling of the size of smallest localized modes: verification of Eq.~(\ref{eq:1}).} \textbf{(a)} Dispersion curves $\omega(\kappa)$ of a W1 waveguide and of a quarter-wave Bragg stack with $n_1=1.5$ and $n_2=3.5$. The former has an effective photon mass that is about 10 times larger than that of the latter. \textbf{(b)} Size threshold $N_\text{th}$ for the two photonic structures with varying $\Delta \omega$ (while keeping $m$ constant for the PhCW) or varying $m$ (while keeping $\Delta \omega$ constant for the Bragg stack). The linearity of the curves validates Eq.~(\ref{eq:1}). The difference in the slopes is likely to be due to the different scattering coefficients at every perturbed interface~\cite{Baron2011}. Furthermore, W1 waveguides exhibit wavelength-scale localized modes, much smaller than those possibly obtained in Bragg stacks, even at tiny disorder levels.}
\label{fig:3}
\end{figure}

We proceed to the numerical verification of the scaling law in Eq.~(\ref{eq:1}). For a careful testing, we consider two different geometries, the W1 waveguide investigated above and a 1D quarter-wave Bragg stack, which have substantially different localization properties. Figure~\ref{fig:3}(a) shows the dispersion curves of the unperturbed photonic structures, where the Bragg stack is composed of alternating dielectric layers with lattice constant $a=453$ nm and refractive indices $n_1=1.5$ and $n_2=3.5$. The band edges of the two photonic structures appear at nearby frequencies, yet the PhCW exhibits an effective photon mass about 10 times larger than the Bragg stack.

The scaling law in Eq.~(\ref{eq:1}) is tested, on the one hand, by varying the disorder level $\sigma$ imposed on the PhCW without changing its structure -- $\Delta \omega$ is therefore varied while $m$ and $\lambda_0$ remain constant -- and, on the other hand, by varying the refractive index $n_2$ of the Bragg stack up to large (unrelatistic for optical waves) values while precisely monitoring the disorder level and the period to respectively maintain $\Delta \omega/c=0.01$ $\mu$m$^{-1}$ and $\lambda_0=1.51$ $\mu$m constant. In the latter case, the sole physical quantity that is expected to vary is therefore the effective photon mass, from $mc \approx$ 4 to 40 $\mu$m$^{-1}$. Furthermore, the same protocol is applied and an average over 10000 independent disorder realizations, obtained by randomly varying the layer thicknesses, is performed (computations rely on simple $2\times 2$ matrix products). The resulting threshold lengths $N_\text{th}$ are shown in Fig.~\ref{fig:3}(b) as a function of $(a^2 m \Delta \omega)^{-1/2}$. A clear linear dependence is obtained in both cases, thereby constituting a firm validation of the scaling law proposed in Eq.~(\ref{eq:1}).

As already noted above, our numerical simulations indicate that wavelength-scale localized modes may be observed in W1 waveguides even at tiny disorder levels. Figure~\ref{fig:3}(b) shows that such small localized modes \textit{cannot} be observed in classical Bragg stacks (except if one considers unrealistically large values for the index contrast). It is the large effective photon mass provided by W1 waveguides that makes the difference.

The formation of localized modes near the photonic band edge is completely expected, but the fact that structural imperfections as small as $\lambda/1000$ may lead to the formation of modes with spatial extents of only a few wavelengths comes as a surprise. In engineered nanocavities, for instance, the lattice structural modifications employed to create wavelength-scale gap modes are usually larger than $\lambda/1000$, typically consisting in removing, shifting or resizing a few holes~\cite{Song2005, Lalanne2008, Notomi2010}. In addition, they are spatially correlated and precisely controlled to collectively contribute to the mode formation. Thus, one would expect rather large volumes for modes created by perturbations that are much weaker and random.

\subsection*{Near-field observation of a wavelength-scale localized mode}\label{sec:experiment}

Our numerical results encouraged us to explore the possibility of observing wavelength-scale localized modes at tiny disorder levels. Towards this aim, we fabricated a W1 waveguide without adding any intentional disorder during the writing process, so that the sole perturbation that remained was the inevitable residual disorder caused by our state-of-the-art fabrication technology. The typical disorder level of the fabrication facility has been characterized by thorough statistical analysis in prior studies and was found to be $\sigma \sim 1-2 \; \text{nm}$ ($\sim 0.001 \lambda_0$)~\cite{Portalupi2011}. Though our procedure lacks control compared to previous studies on localization that introduce intentional random perturbations~\cite{Topolancik2007, Sapienza2010, Gao2013, Thyrrestrup2012}, we are able to reach a minute and unprecedented level of perturbation that has never been probed up to now near photonic band edges. 

\begin{figure}[t]
\centering
\includegraphics[width=\linewidth]{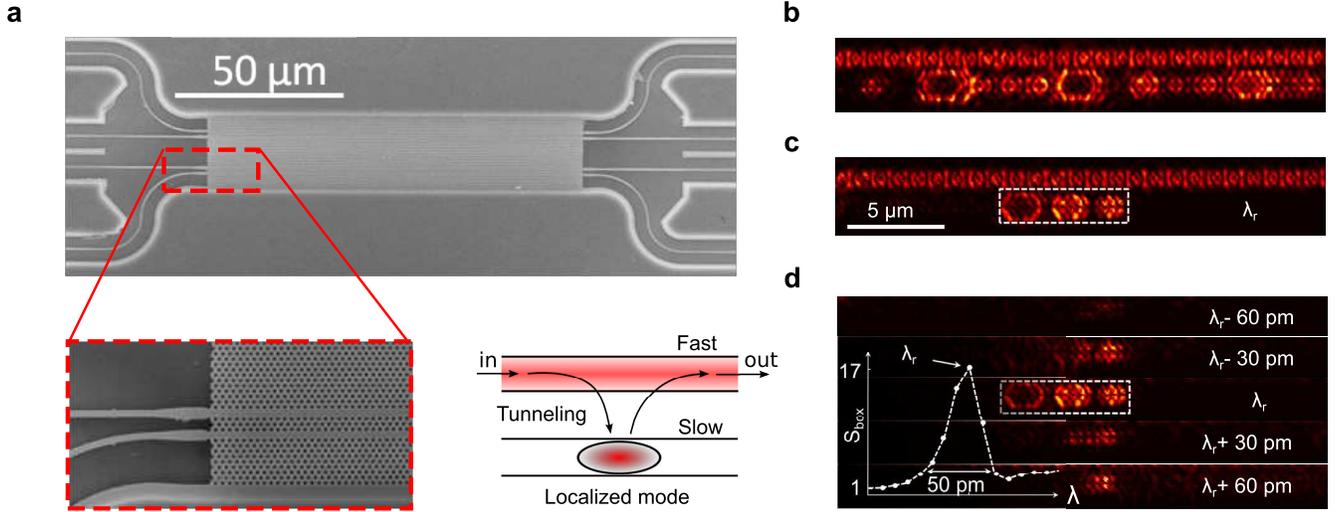}
\caption{\textbf{Near-field experiment.} \textbf{(a)} Scanning electron microscope images of the sample. The inset on the bottom left provides a close-up of the two side-coupled 84-$\mu$m-long PhCWs and of the access ridge waveguides used to inject light. The inset on the bottom right shows a sketch of the layout designed to obtain clear near-field images of the localized modes. Light is injected in the W1.1 waveguide, which operates in the fast-light regime, and couples evanescently to localized modes in the W1 waveguide, operating in the slow-light regime. \textbf{(b,c)} High-resolution (10 pm) near-field images of the PhCW pair recorded over an area covering the first 25 $\mu$m of the sample. The panel in \textbf{(b)} shows an extended state ($\lambda=1489.61$ nm) composed of a series of coupled localized modes. The panel in \textbf{(c)} shows a wavelength-scale localized mode ($\lambda=1488.38$ nm) composed of three sub-spots and of spatial extent about 6 $\mu$m. It is the smallest localized mode detected experimentally. \textbf{(d)} Spectral evolution of the wavelength-scale mode near the resonance wavelength $\lambda_r$. The inset shows the intensity $S_\text{box}(\lambda)$ integrated over the rectangular dashed-linebox and normalized to the averaged intensity in the W1.1 waveguide.}
\label{fig:4}
\end{figure}

At vanishingly small group velocities, it is well known that the optical mean free-path is very short and coupling light into localized modes cannot be achieved by end-fire injection through cleaved facets~\cite{Notomi2001}. For this reason, we designed a layout consisting of a pair of collinear waveguides: a W1 waveguide operating in the slow-light regime near the band cutoff-wavelength $\lambda_0 \sim 1490$ nm, and a W1.1 (10\% larger defect-width) waveguide operating in the fast-light regime, and therefore much less sensitive to residual imperfections. An SEM micrograph of a typical set of waveguides fabricated into a 220-nm thick free-standing silicon membrane is shown in Fig.~\ref{fig:4}(a). The fast (W1.1) waveguide is used as an independent channel for delivering light into the localized modes supported by the slow (W1) waveguide via evanescent coupling, see the inset of Fig.~\ref{fig:4}(a). This coupling is extremely weak due to the very low disorder level and is therefore expected not to affect the localization properties of interest here. Furthermore, compared to a previous approach where TE-like localized modes were excited via a weak coupling with TM-like ballistic modes in a single PhCW, thereby resulting in near-field images containing both localized and extended modes~\cite{Huisman2012a}, our configuration allows us to form clear near-field maps of individual localized modes and thus, estimate their spatial extent with greater accuracy.

Nevertheless, because of their small spatial extents and spectral bandwidths as well as their unknown spatial and spectral positions, wavelength-scale localized modes remain challenging to observe. Hence, for a rapid initial exploration over broad spatial and spectral intervals, we first use low-spectral ($200$ pm) and low-spatial resolution ($100$ nm) multispectral near-field scanning-optical microscopy (SNOM) using a tapered monomode silica fiber probe and a supercontinuum laser source~\cite{Dellinger2012}. Only the first 25 $\mu$m of the waveguide were probed. Due to the lack of spectral resolution, the images recorded in the $1480-1500$ nm band and shown in the Supplementary Movie 1 online cannot reveal localized modes with a quality factor $Q > 10^4$. We distinguish two types of states for wavelengths close to the band edge of the W1 waveguide at $1487$ nm. The most frequent ones are delocalized states that extend over most of the scanned area and exhibit numerous spatial bright spots. More interesting in the present context is the existence of less probable, wavelength-scale localized modes: two modes at $\lambda=1488.3$ and $1491.5$ nm are seen in the movie. With high-resolution SEM analysis of the W1 waveguide, we have checked that they are not due to abnormally large, albeit inevitable, local fabrication defects, but rather to the residual intrinsic imperfections inherent to the fabrication process over the entire waveguide structure, see the Supplementary Information online.

For a deeper analysis, we resort to high spectral-resolution SNOM with a $1$-pm-resolution tunable laser, increasing the spatial sampling to $\sim 62$ nm and scanning the W1 waveguide at spatial locations found with low resolution. This allows us to detect modes with larger $Q$'s, but in return, finding a resonance becomes very time-consuming. Figure~\ref{fig:4}(b) shows a state formed by a chain of spots with varying brightness, which collectively resonate at the same wavelength, ergo all spots including the less intense spots belong to the same coherent state that covers the entire scan interval. The ``stadium-shaped'' patterns are due to the cavity tip interaction~\cite{Mujumdar2007}. Figure~\ref{fig:4}(c) shows a different mode composed of three dominant main spots that are grouped together and show up over a dark background. It is the smallest mode that we have observed. Its total spatial extent $L$ is smaller than $6$ $\mu$m. For comparison, the numerical data reported in Fig.~\ref{fig:2}(d) predict that $P_{1.5}(L_0)$ is equal to 0.14 for $L_0=6$ $\mu$m. This implies that, on average along the W1 waveguide, localized modes with spatial extents $L \leq 6$ $\mu$m are expected every $L_0/P_{1.5}(L_0) = 43$ $\mu$m. This is consistent with our observation of a single localized mode obtained by scanning a 25-$\mu$m-long section of the W1 waveguide. Interestingly also, we note that the predicted and measured lower bound in mode spatial extent matches well with the cavity length above which the operation of photonic-crystal cavity lasers becomes significantly impacted by residual fabrication imperfections~\cite{Xue2016}. The occurrence of disorder-induced localization in engineered cavities is an aspect that deserves more attention.

Figure~\ref{fig:4}(d) finally shows the spectral evolution of the localized mode, calculated by integrating the measured intensity over a fixed rectangular area comprising the mode and normalizing it to the averaged intensity in the W1.1 waveguide for several wavelengths. The result confirms the existence of a high confinement level both in the spatial and spectral domains. As the wavelength is tuned away from resonance at $\lambda=1488.38$ nm, we observe that the spatial and spectral variations of the spot intensities exhibit an intricate behavior, suggesting a beating between several modes (although non-uniform coupling with the tip cannot be excluded). The inset shows the spectral evolution of the normalized intensity of the localized state and evidences a resonance with a $Q \sim 5.10^4$. Since the localized mode is formed from tiny structural modifications, smaller than those typically employed for engineered cavities, mode-profile impedance mismatch~\cite{Lalanne2008} is kept at a very low level, and leakage into the air cladding and into the W1.1 waveguide is expected to be as small as that encountered with side-coupled engineered cavities. Thus, we believe that the observed $Q$ value is limited by the tip interaction, consistently with earlier works with silica tips and engineered cavities~\cite{Koenderink2005}.

\section*{Discussion}\label{sec:discussion}

In this article, we have investigated the physical mechanism underlying the formation of small localized modes at band edges of periodic media, demonstrating the existence of a minimal mode size and showing that this bound is predominantly driven by the effective photon mass. In particular, we have found that wavelength-scale localized modes naturally form up in PhCWs at state-of-the-art intrinsic disorder levels due to the flatness of the dispersion curve.

The localized modes have positions and frequencies that are not known in advance by design. As such, they are not easy to handle in applications that require extreme precisions, but their existence and the possibility to enhance (statistically) their spatial confinement by using flatter dispersion curves is extremely relevant for quantum electrodynamics experiments~\cite{Sapienza2010, Thyrrestrup2012, Gao2013, Minkov2013}, random lasing~\cite{Yang2011a, Liu2014}, but also for sensing applications~\cite{Wang2010a, Scullion2013} and random photonics devices~\cite{Vynck2012, Redding2013}.

The existence of a lower bound for the volume of localized states in perturbed periodic media and the importance of the effective photon mass have not been pointed out in early works on optical localization in the slow-light regime. These first results, supported by a simple intuitive model, should encourage more in-depth theoretical investigations.

Similarly, the effective photon mass is rarely acknowledged in the literature on engineered nanocavities, such as the heterostructure family~\cite{Song2005, Kuramochi2006}, which are defect-modes that benefit from a slow-light effect close to the band edge~\cite{Lalanne2008} and, as explained in this article, are in many respects similar to the present localized modes. We therefore expect that the conceptual understanding gained from this disorder-driven study will have repercussions on future photonic structure designs to push back the ultimate limit imposed by unavoidable disorder in slow-light photonic devices~\cite{Hughes2005, Mazoyer2009, Sancho2012} as well as in engineered cavities~\cite{Taguchi2011, Xue2016}.

\section*{Methods}

\subsection*{Numerical simulations and protocol}

All our computational results were obtained with an in-house fully-vectorial frequency-domain Fourier-Bloch-mode method~\cite{Silberstein2001, Lecamp2007b}. The statistical retrieval of $P_\sigma(L_0)$ requires determining whether the resonant modes identified in the LDOS spectrum are truly individual localized modes. On the one hand, the field profile of localized modes (with spatial extent smaller than the computational system size) should remain unchanged after increasing the waveguide length. On the other hand, a normalized field profile independent of the position of the source indicates that a single mode contributes to it. On this basis, we computed the on-axis magnetic-field profiles generated by a dipole source at the center of the system at the resonance wavelength for the 100$a$-long perturbed waveguide, for a 200$a$-long protracted waveguide obtained by surrounding the 100$a$-long waveguide by two 50$a$-long perturbed sections, and for the same 200$a$-long waveguide where the source position was shifted by two periods. A resonant mode was considered as an individual localized mode only if the field profile remained unchanged after the waveguide extension and source displacement. The spatial extent of the localized mode is defined as $L=|z_1-z_2|$, where the magnetic field intensity $|H(z)|^2$ should be smaller than $\text{max}(|H(z)|^2)/e^2$ everywhere outside the interval $[z_2; z_1]$.

\subsection*{Sample fabrication}

The PhCWs were defined in a ZEP-520A resist by electron beam lithography, using a 30kV Raith/LEO lithography system. The pattern was transferred into the 220 nm top layer of silicon-on-insulator wafers by a SF$_6$/CHF$_3$ reactive ion etch. After stripping of any remaining ZEP resist, the buried oxide under the photonic crystal region was removed by selective wet etching with HF acid. Finally, samples were cleaned in a H$_2$O$_2$/H$_2$SO$_4$ mixture and cleaved to allow butt coupling. The fabrication procedure is explained in more detail in~\cite{Reardon2012}. The structural parameters of the fabricated sample vary slightly from those used in the simulations, yielding a photonic band edge at about $1490$ nm, which is clearly observed in transmission measurements (see the Supplementary Information online).

\subsection*{Near-field measurements}

Near-field measurements were made using a home-made Scanning Near-field Optical Microscope (SNOM) operating in collection mode with a shear-force feedback. A pulled silica fiber with a 50 nm apex was used as a near-field probe. The sample-probe distance was set to 10 nm. A butt coupling technique with microscope objectives was employed for light injection inside the integrated waveguides and light polarization was controlled with a free-space polarizer.

\section*{Acknowledgements}

PL thanks Jean-Paul Hugonin and Simon Mazoyer for fruitful discussions. RF has received financial support from the French ``Direction Générale de l’Armement'' (DGA). Near-field experiments by LL, BC and FF were made in the framework of the LABEX ACTION (ANR-11-LABX-0001-01).

\section*{Author contributions statement}

R.F. and X.Z. performed the numerical calculations. A.B., P.L. and S.A.S. developed the device design. S.A.S. fabricated the sample. A.B. performed the end-fire transmission measurements. L.L. performed the near-field experiments. P.L. and K.V. supervised the theoretical investigations. T.F.K. supervised the device design and fabrication. B.C. and F.F. guided the near-field investigations. P.L. and A.B. supervised the overall project. P.L., K.V., A.B., R.F. and T.F.K. wrote the manuscript. All authors contributed to the analysis of the results and corrected the manuscript.

\section*{Additional information}

\textbf{Competing financial interests}
The authors declare no competing financial interests.


\part*{Supplementary Information}

\section*{Numerical results}

\subsection*{Computational method}
A systematic exploration of real PhCWs with long-scale propagation lengths and nanometer perturbations is out of reach of present state-of-the-art 3D computational approaches. For this reason, we resort to a 2D fully-vectorial analysis with an effective index of 2.83 for the guided mode of a 220-nm-thick silicon slab in air. Out-of-plane scattering into the air cladding is therefore omitted in the computation, but since this loss channel is much weaker than the backscattering channel near the band edge, the approximation is likely to impact only weakly our predictions on the spatial extent of the cavity modes.

All our computational results are obtained with an in-house fully-vectorial frequency-domain Fourier-Bloch-mode method~\cite{Silberstein2001, Lecamp2007b}. The strength of this method is that it relies on an analytical integration of Maxwell’s equations along the PhCW axis, allowing us to study long waveguides with an S-matrix formalism which can handle Bloch modes~\cite{Lecamp2007b}. Another unique feature of the approach is its capability to analytically satisfy the outgoing Bloch-wave conditions at the PhCW termination. This provides a high degree of accuracy to the computational results, even near the band edge frequency. We resort to numerical integration only in the transverse x-direction with a super cell approach and Perfectly-Matched-Layers~\cite{Lecamp2007b}. The method has already been used for many studies of periodic waveguides and has been compared to experimental investigations of real situations involving PhCWs. For instance, it has been used to study various effects related to the emission of dipole sources in PhCWs~\cite{Lecamp2007} and to the transport of light in disordered PhCWs~\cite{Mazoyer2010, Baron2011}.

\subsection*{Individual localized modes}

\begin{figure}[ht]
\centering
\includegraphics[width=0.95\linewidth]{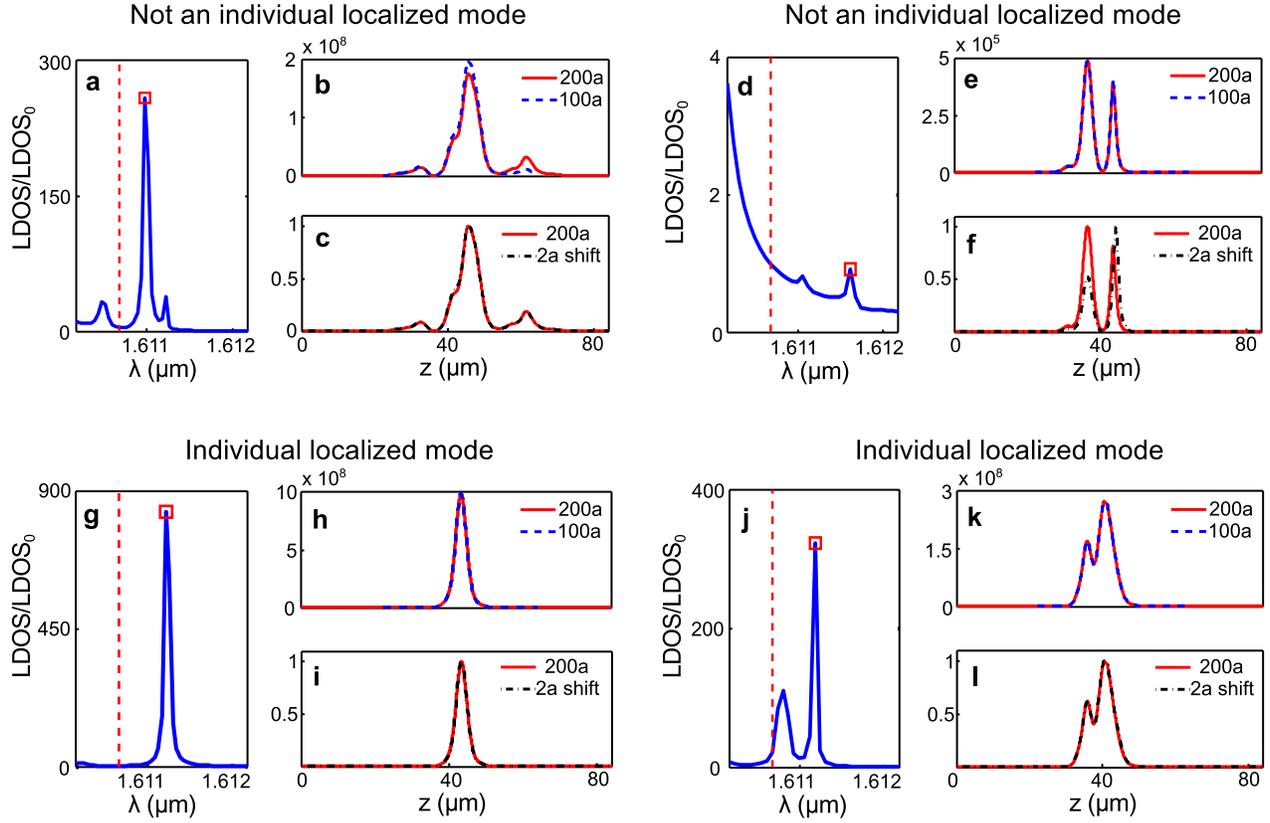}
\caption{\textbf{Examples of resonances that are or are not accounted for in the distribution function of spatial extent of localized modes.} \textbf{(a,d,g,j)} Spectral dependence of the normalized LDOS obtained for a source placed in the center of the $100a$-long W1 waveguide. The vertical red dashed line indicates the band-edge wavelength $\lambda_0$. \textbf{(b,e,h,k)} Envelop of the magnetic-field-intensity profiles, $|H_{100}|^2$ and $|H_{200}|^2$, at the resonance wavelength for the $100a$-long W1 waveguide (dashed blue line) and the extended $200a$-long W1 waveguide (red solid line). \textbf{(c,f,i,l)} Envelop of the magnetic-field-intensity profiles in the $200a$-long W1 waveguide for a source positioned in the center (red solid line) and for the same right-shifted source (black dashed-dotted line). \textbf{(a-c)} Resonance that is \textit{not} considered as an individual localized mode (mode profile affected by the boundary). Here, $\sigma=0.5$ nm. \textbf{(d-f)} Resonance that is \textit{not} considered as an individual localized mode (mode profile affected by the source position). Here, $\sigma=0.75$ nm. \textbf{(g-i)} Resonance that is considered as an individual localized mode. The extension length is $L=6.4$ $\mu$m and $\sigma=0.5$ nm. \textbf{(j-l)} Resonance that is considered as an individual localized mode. The extension length is $L=12.5$ $\mu$m and $\sigma=0.75$ nm.}
\label{fig:s1}
\end{figure}

The protocol used to determine whether the resonances observed in the local density of states (LDOS) spectrum correspond to individual localized modes is described in the Methods section. In brief, the field profile of individual localized modes (with spatial extent smaller than the system size) should remain unchanged after extending the waveguide from 100$a$ to 200$a$ and their normalized field profile be independent of the source position. Figure~\ref{fig:s1} shows several examples of resonant modes corresponding or not to individual localized modes. Note that individual localized modes are retrieved independently of the physical mechanism underlying their formation (Anderson localization or gap confinement)~\cite{Bliokh2008} and all delocalized modes, including necklace states~\cite{Bertolotti2005, Sebbah2006}, are excluded.

\section*{Characterization of the fabricated sample}

\subsection*{Transmission measurements}

\begin{figure}[ht]
\centering
\includegraphics[width=0.95\linewidth]{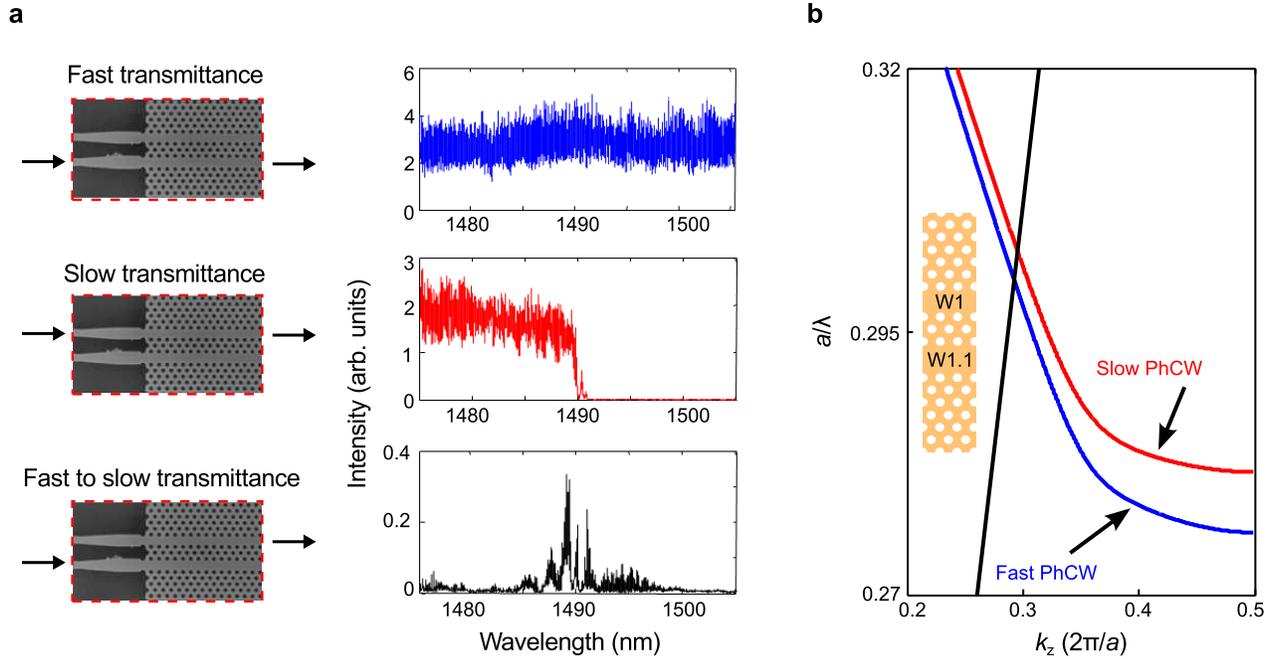}
\caption{\textbf{End-fire characterization of the side-coupled PhCWs.} \textbf{(a)} Transmission spectra of the W1.1 and W1 waveguides (top blue and middle red curves respectively) when light is injected from the access ridge waveguides. The bottom black curve is the spectrum of the signal collected at the output of the W1 waveguide, when light is injected into the W1.1 waveguide. \textbf{(b)} Calculated dispersion curves of the W1 (red) and W1.1 (blue) waveguide for the experimental structural parameters.}
\label{fig:s2}
\end{figure}

Prior to near-field characterization, the fabricated sample was characterized via spectroscopic transmission measurements performed at telecommunication wavelengths. As shown in the inset of Fig.~\ref{fig:s2}(a), both the slow (W1) and fast (W1.1) waveguides have access ridge waveguides in order to couple light into the structure. Note also that the first five periods of the waveguides are modified to efficiently couple light into the PhCW mode, independent of the group index. The W1 and W1.1 waveguide transmission spectra were recorded with a high-resolution tunable external laser source ($1450-1620$ nm) using an end-fire setup. The measured spectra are shown as blue and red curves in Fig.~\ref{fig:s2}(a). As expected from the calculated dispersion curves in Fig.~\ref{fig:s2}(b), the W1 transmission drops abruptly near the band-edge wavelength ($\lambda_0 \sim 1490$ nm), while the fast W1.1 waveguide remains transparent for the input signal way beyond $\lambda_0$. The bottom black curve corresponds to the transmission spectrum recorded at the W1 waveguide output when light is injected into the fast W1.1 waveguide. The amount of transmitted light that we observe in the W1 band gap for $\lambda>\lambda_0$ is due to the coupling between the two waveguides. It evidences the possibility of energy transport in the forbidden band of the W1 waveguide over distances that are smaller than the 200-$\mu$m waveguide length.

\subsection*{High-resolution scanning electron microscope analysis}

\begin{figure}[htbp]
\centering
\includegraphics[width=0.5\linewidth]{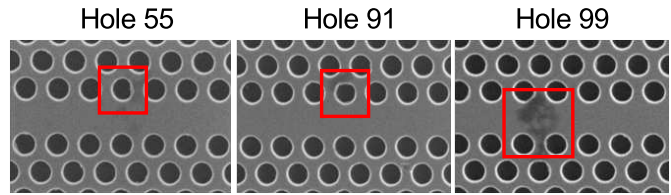}
\caption{\textbf{High-resolution scanning electron microscope images of the three major defects in the W1 waveguide exhibiting a localized mode.} These imperfections are located at holes number 55, 91 and 99, i.e. at 23.52 $\mu$m, 38.22 $\mu$m and 41.58 $\mu$m respectively from the input facet of the W1 waveguide. Holes 55 and 91 exhibit resist stains and visible impurities around the hole. Hole 99 exhibits a resist stain and severe distortion of the hole.}
\label{fig:s3}
\end{figure}

To assert that the wavelength-scale localized mode reported in the main text is indeed due to the residual imperfections left by our fabrication facility and not to abnormally large imperfections resulting from a failure of the fabrication process, we carefully analyzed the W1 waveguide that produced the localized mode under a high-resolution scanning electron microscope (SEM). Figure~\ref{fig:s3} shows the three major defects observed in the W1 waveguide which consist in resist stains and small hole deformations. These defects are not observed at the locations of the observed localized mode under the near-field optical microscope shown in the main text. We can thus infer that the formation of localized modes are due to residual imperfections. This conjecture is supported by the computational results that predict that wavelength-scale localized modes exist for disorder levels that are even smaller than those of our fabrication process.

\end{document}